\documentstyle[11pt,example-proceeding,epsfig]{article}

\begin{document}
\vspace*{4cm}
\title{Four fermion processes at LEP II}

\author{ V. Verzi }

\address{Dipartimento di Fisica, Universit\`a di Milano and INFN-MILANO, Via Celoria 16,  \\
IT-20133 Milan, Italy.}

\maketitle\abstracts{
In its second phase, LEP has given the unique opportunity to study four fermion processes never observed before.
This papers deals with a preliminary study of $WW$, $We\nu$, $ZZ$ and $Z\gamma^\star$ production in $e^+e^-$ collisions.
Measurement results on $WW$ cross section, $W$ branching fractions, $WWV$ ($V=Z/\gamma$)  
anomalous couplings and $W$ average polarization in $WW$ events 
are presented. Moreover cross section measurements for $We\nu$, $ZZ$ and $Z\gamma^\star$ production
and limits on $ZZZ$ and $ZZ\gamma$ anomalous couplings are reported. 
All the results are in good agreement with the Standard Model expectations.}

\vspace*{6cm}
\begin{center}
{\small Talk presented in\\
XXXVI Recontres de Moriond Electroweak Interactions and Unified Theories, 10-17 March 2001.}
\end{center}

\newpage
\section{Introduction}

In its second phase, LEP has been operated in years from 1996 to 2000 at centre-of-mass 
energies ($\sqrt{s}$) between 161~GeV and 208~GeV. This has allowed to each of the four experiments
to collect nearly 700~$pb^{-1}$ of data above the $W$-pair production threshold, as summarised in table 
\ref{tab:lum}.
\vspace*{-0.4cm}
\begin{table}[h]
\caption{Approximated value of the integrated luminosity collected by the DELPHI experiment at LEP II.}
\vspace{0.2cm}
\begin{center}
\begin{tabular}{|l|l|l|l|l|l|l|l|l|l|l|l|}
\hline
$\sqrt{s}$ (GeV)            & 161    & 172    & 183    & 189     & 192    & 196    & 200    & 202    & 205    & 207     \\ \hline
$\int L dt\ (pb^{-1})$      &$\sim$10&$\sim$10&$\sim$55&$\sim$160&$\sim$25&$\sim$75&$\sim$85&$\sim$40&$\sim$85&$\sim$140\\ \hline
year                        & 1996 & 1996 & 1997 & 1998 & 1999 & 1999 & 1999 & 1999 & 2000 & 2000 \\ \hline
\end{tabular}
\end{center}
\label{tab:lum}
\vspace*{-0.1cm} 
\end{table}

The high centre-of-mass energy achieved at LEP II has allowed to study several four fermion processes  
and hence to test the Standard Model (SM) of electroweak interactions in sectors only poorly known before.
The most important process is the $WW$ production because it allows to measure the $W$ mass \cite{m01:mw}
and the triple gauge boson couplings $WWV$ ($V=Z/\gamma$) (TGC) which appear
at the tree level.
Other four fermion final state are attained with the production of a pair of neutral gauge bosons $ZZ$ and $Z\gamma^\star$.
These processes may receive New Physics contributions from triple neutral gauge boson couplings $ZZZ$, $ZZ\gamma$ and 
$Z\gamma\gamma$ which are expected to be unobservably small in the SM. 
The remaining four fermion processes involve the production of a single gauge boson: $We\nu_e$ and $Zee$. 
Particularly important is $We\nu_e$ because it allows to further constrain the $WW\gamma$ vertex \cite{yellow:lep2-1}.

This paper presents the results on four fermion processes with particular attention to the preliminary measurements
obtained with the data collected in the year 2000.
The second section deals with $WW$ and $We\nu$ processes while the third deals with $ZZ$ and $Z\gamma^\star$ production.
No new preliminary results on $Zee$ process have been presented and the old ones are described elsewhere \cite{zee-osaka}.

\section{$WW$ and $We\nu$ production}

With the full luminosity available at LEP, about 11000 $WW$ events are expected to be produced 
in each of the four experiments. According to the $W$ decay modes, one has three different topologies in 
$WW$ events: fully hadronic ($WW \rightarrow qqqq$), semileptonic ($WW \rightarrow qql\nu_l$) and
fully leptonic ($WW \rightarrow l \nu_l l \nu_l$) events respectively with branching fraction of 
46\%, 43\% and 11\%. These events can be selected with high efficiencies 
and high purities.

\begin{figure}
\centerline{\epsfig{file=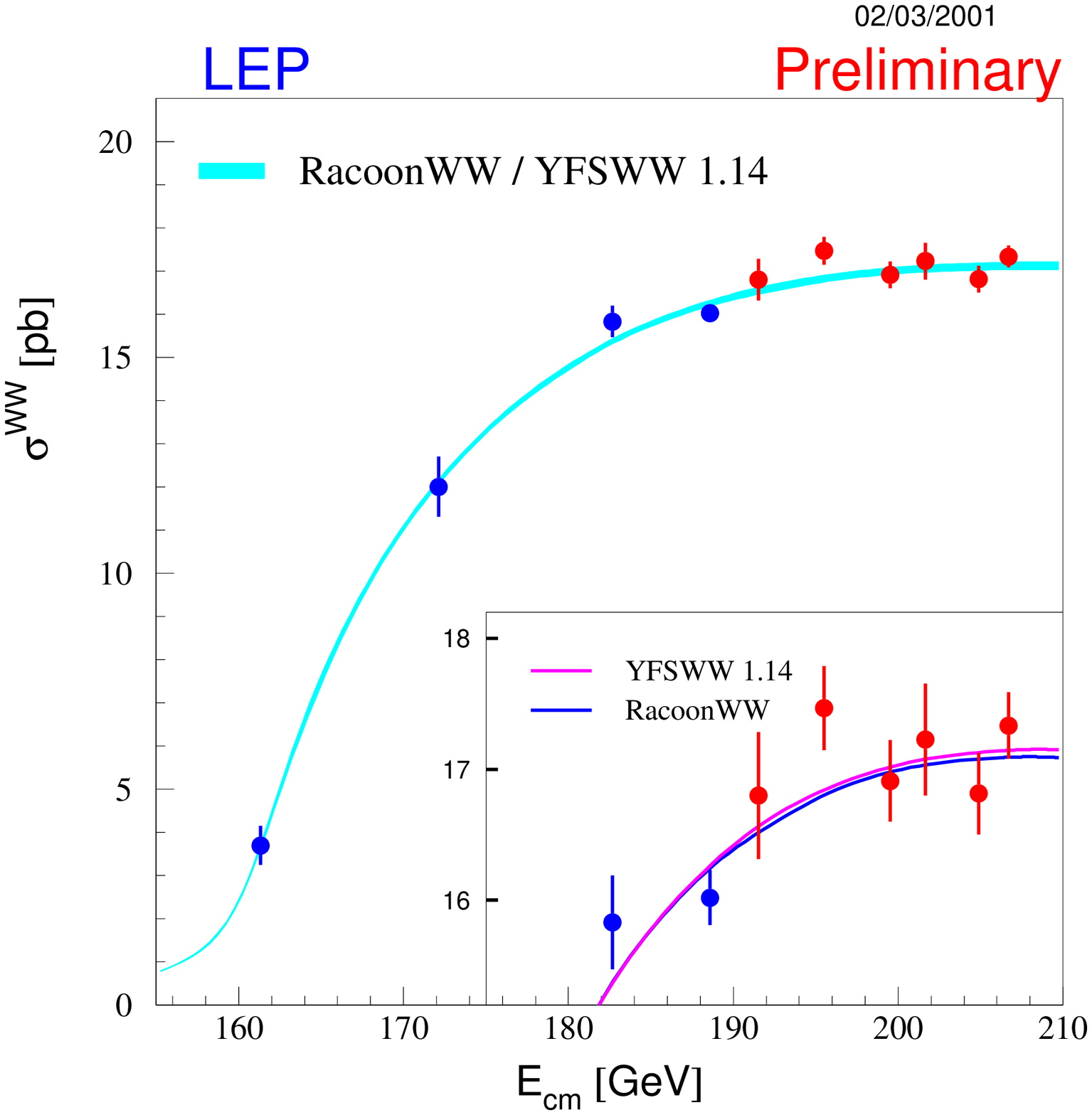,width=6.0cm}\hfill\epsfig{file=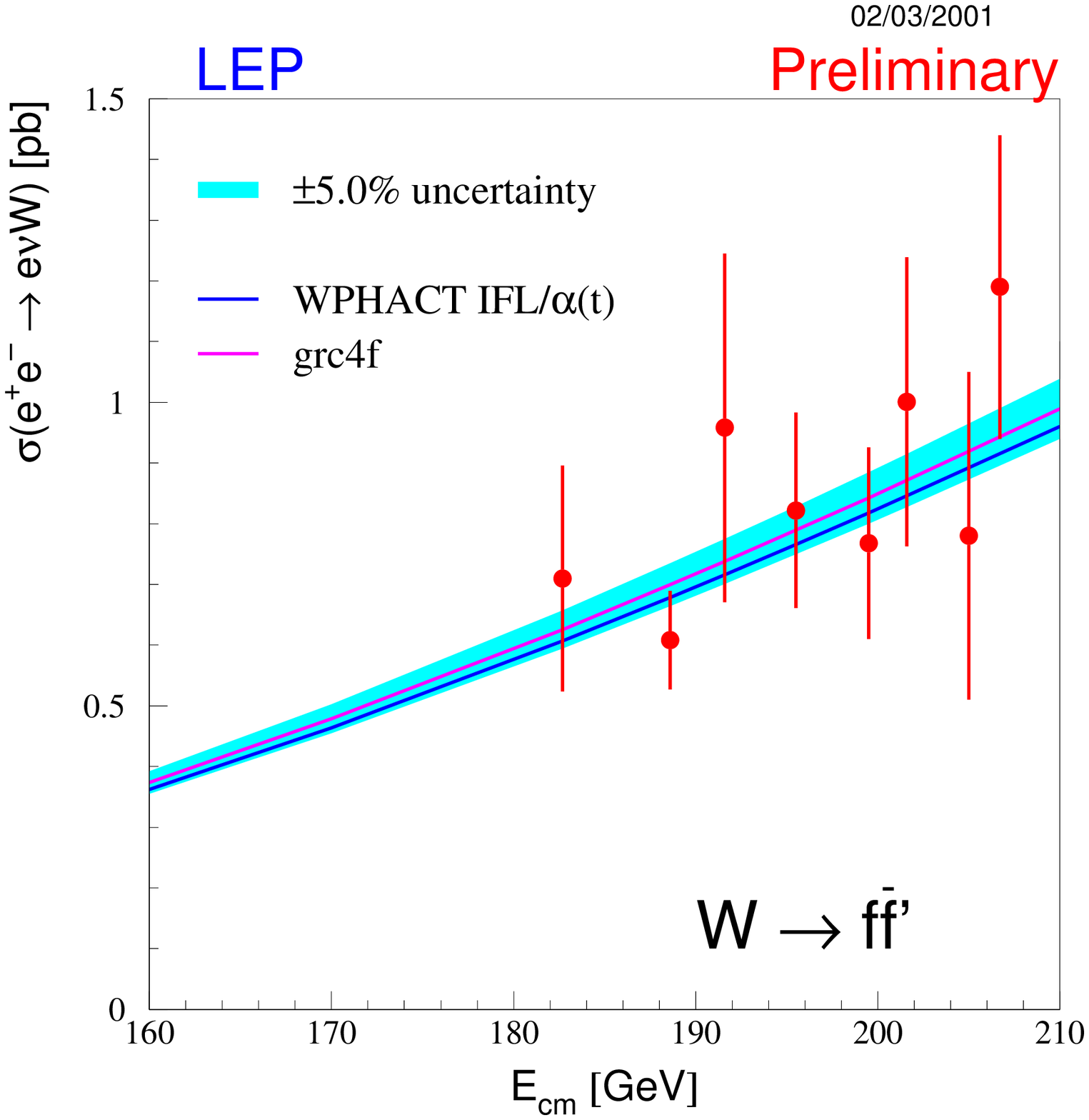,width=6.0cm}} 
  \vspace*{-7.0cm} 
\centerline{\epsfig{file=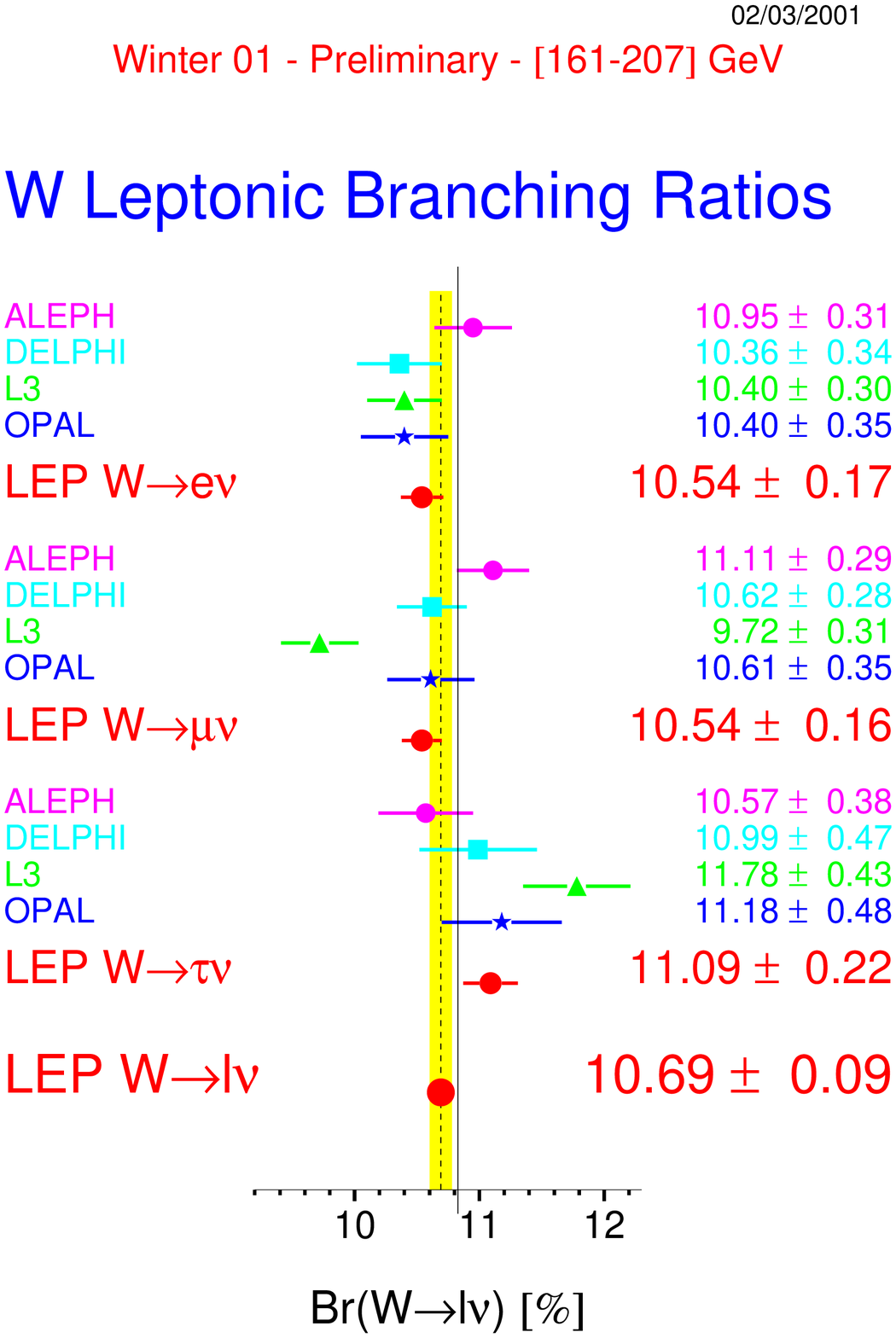,width=6.0cm}} 
  \vspace*{-1.5cm}
  \caption{LEP combined results$^7$
           on $WW$ cross section (on the left), on $W$ branching fractions 
          (central figure) and on $We\nu$ cross section (on the right). 
          The leptonic branching fraction, $Br(W\rightarrow l\nu) = (10.69 \pm 0.09)\%$,
          is measured assuming the lepton universality, and one 
          can derive the hadronic branching fraction, 
          $Br(W\rightarrow qq) = 1 - 3 Br(W\rightarrow l\nu) = (67.92 \pm 0.27)\%$, 
          in agreement with the SM value $Br(W\rightarrow qq)|_{SM} = 67.51 \%$.
          }
  \label{fig:csww-br-cswen} 
\vspace*{-0.1cm} 
\end{figure}

The four experiments have measured the $WW$ cross section at all energies above 161~GeV \cite{csww:00}. 
The combined results are shown in figure \ref{fig:csww-br-cswen}, compared with the theoretical 
predictions of RacoonWW \cite{racoonww} and YFSWW \cite{yfsww} programs that treat the ${\cal O}(\alpha)$ radiative 
corrections in ``Double Pole Approximation'' (DPA). The theoretical error for DPA is $\le 0.5\%$
and it is small compared to the error of 2\% wich characterize the calculation of ${\cal O}(\alpha)$ corrections 
in ``Improved Born Approximation'' (IBA) \cite{yellow:lep2-1}. 

The $W$ decay branching fractions 
$Br(W \rightarrow ff)$ have been determined from the cross section for the individual $WW \rightarrow 4f$ decay channels, 
and the results are shown in figure \ref{fig:csww-br-cswen}. The  
measurement of $B(W \rightarrow qq)$ allows an indirect determination \cite{ewg:cs} 
of the Cabibbo-Kobayashy-Maskawa matrix element $|V_{cs}|$: $|V_{cs}|_{LEP} = 0.996 \pm 0.013$.

The most general Lorentz invariant Lagrangian for $WWV$ ($V=Z/\gamma$) vertex interactions, under some 
reasonable requirements \cite{yellow:lep2-1}, can be described by three independent couplings 
$\{ \Delta g_1^Z, \Delta k_\gamma, \lambda_\gamma \}$ \footnote{$\Delta c = c - 1$ indicates
the deviation from the SM expectation for $c$ that is 1 while in SM $\lambda_\gamma = 0$.} (TGC)
and another two constrained couplings, 
$\Delta k_Z = \Delta g_1^Z + \Delta k_\gamma \tan^2\theta_W$ and $\lambda_Z = \lambda_\gamma$. 

The TGCs are measured using the information from the total cross section and from the shape of the 
distributions for physical observables. Due to the high precision achieved by LEP experiments, 
these distributions should be predicted including the ${\cal O}(\alpha)$ corrections in DPA. 
However, the generator programs used by the LEP experiments, wich include the description of the detectors, 
treat the ${\cal O}(\alpha)$ corrections using the IBA. Only the ALEPH collaboration \cite{tgc-aleph} has estimated the 
systematic effect due to the missing DPA, showing that it contributes in a sizeable 
way to the total systematic error for the couplings $\Delta g_1^Z$ and $\lambda_\gamma$. 
ALEPH is the only collaboration that has updated the results \cite{tgc-aleph} on TGCs including the 
data collected in 2000 obtaining the following 95\% confidence intervals: 
$[-0.048,+0.080]$ for $\Delta g_1^Z$, 
$[-0.164,+0.134]$ for $\Delta k_\gamma$ and $[-0.059,+0.065]$ for $\lambda_\gamma$. 
No new LEP combined values have been produced after the 2000 summer conferences. It's wortwhile to note 
that the missing DPA should not change drammatically the LEP combined results \cite{lwg:tgc} because for 
$\Delta g_1^Z$ and $\lambda_\gamma$ the statistical error dominates.

A model independent way to test the SM in the $WW$ production consists in comparing the measured average $W$ polarization 
$f_\lambda$ ($\lambda$ indicates the $W$ polarization) with the theoretical expectations. 
The L3 collaboration has done this measurement in the semi-leptonic channel \cite{wpol-l3}. Using the data collected 
in 2000, they have measured $f_0 = (21.6 \pm 5.3)\%$,  $f_{-1} = (64.7 \pm 6.6)\%$ and $f_{+1} = (13.7 \pm 3.4)\%$ that are 
in agreement with the SM values: $f_0^{SM} = 22.0 \%$, $f_{-1}^{SM} = 62.3 \%$ and $f_{+1}^{SM} = 15.7 \%$.

The LEP combined measurements results for $We\nu$ cross section are shown in figure \ref{fig:csww-br-cswen}. Since the 2000 
winter conferences, the only new preliminary results have been presented by the ALEPH collaboration \cite{wen-aleph} 
which has analysed the data collected in the year 2000.

\section{$ZZ$ and $Z\gamma^\star$ production}

According to the $Z$ decay modes there are five visible decay channels. The dominant ones are the 
four jets channel ($ZZ \rightarrow qqqq$), the two jets plus missing energy channel
($ZZ \rightarrow qq\nu\nu$) and the two jets plus two leptons 
($ZZ \rightarrow qqll$) channel with an expected branching ratio of 49\%, 28\% and 14\% respectively.

\begin{figure}
\centerline{ \hspace*{0.8cm} {\scriptsize 4-jets DELPHI LEP2 (Preliminary)} \hfill ${\scriptscriptstyle e^+e^- \rightarrow q\bar{q}\mu^+\mu^-}$ \hspace*{2.5cm}}
\vspace*{-0.5cm} 
\centerline{ \hspace*{0.6cm} \epsfig{file=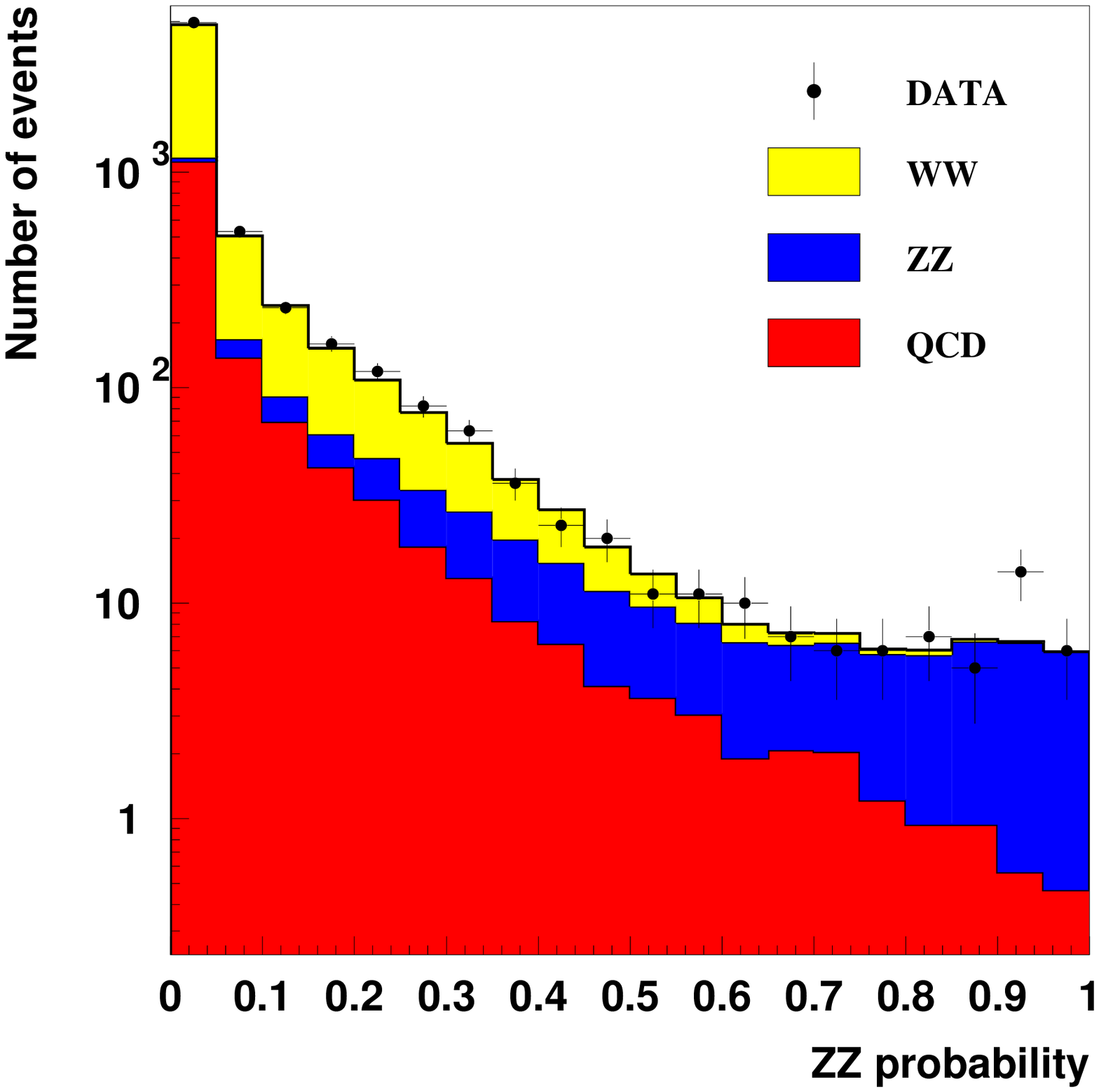,width=6.0cm} 
             \hspace*{-0.8cm} \epsfig{file=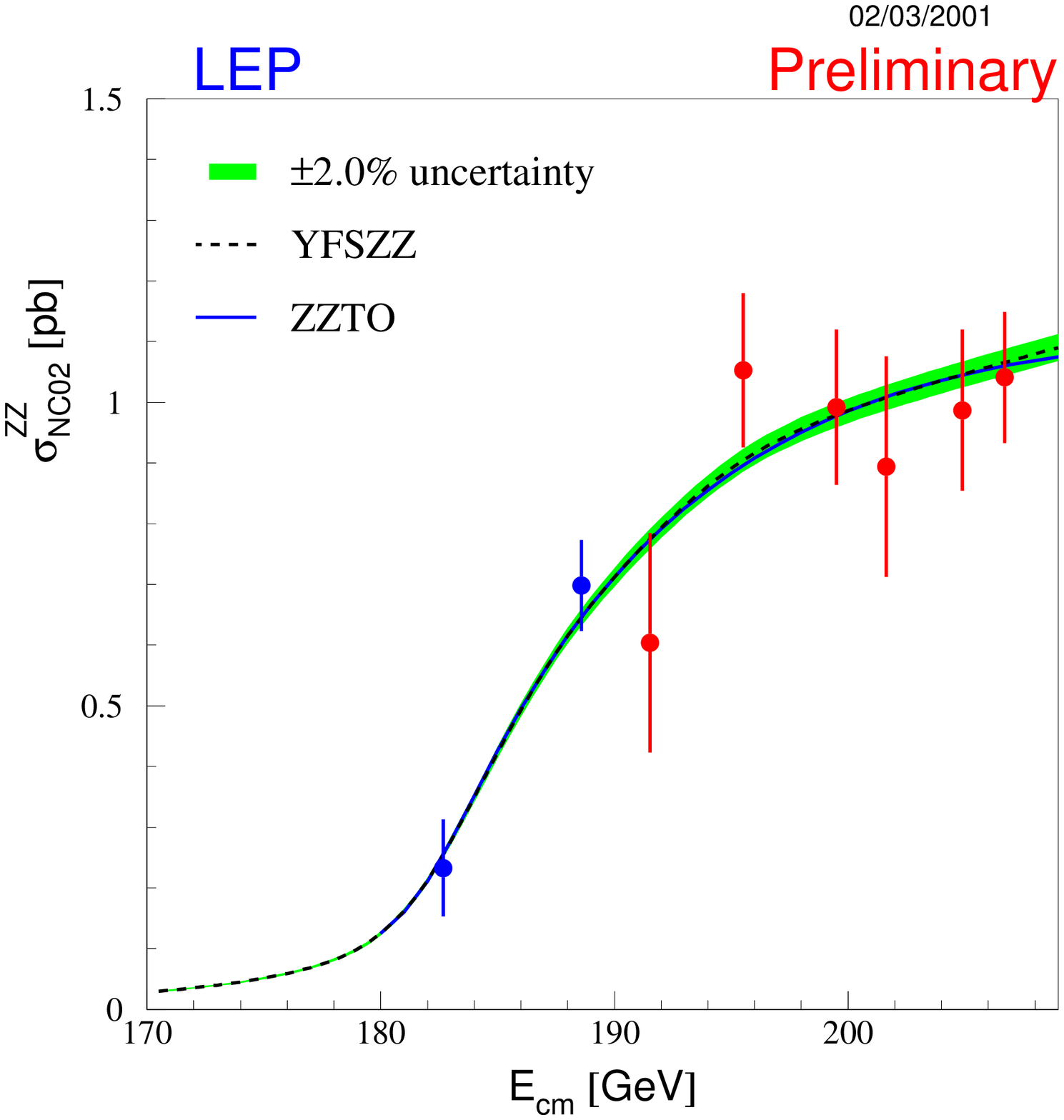,width=6.0cm} 
             \hspace*{-0.5cm} \epsfig{file=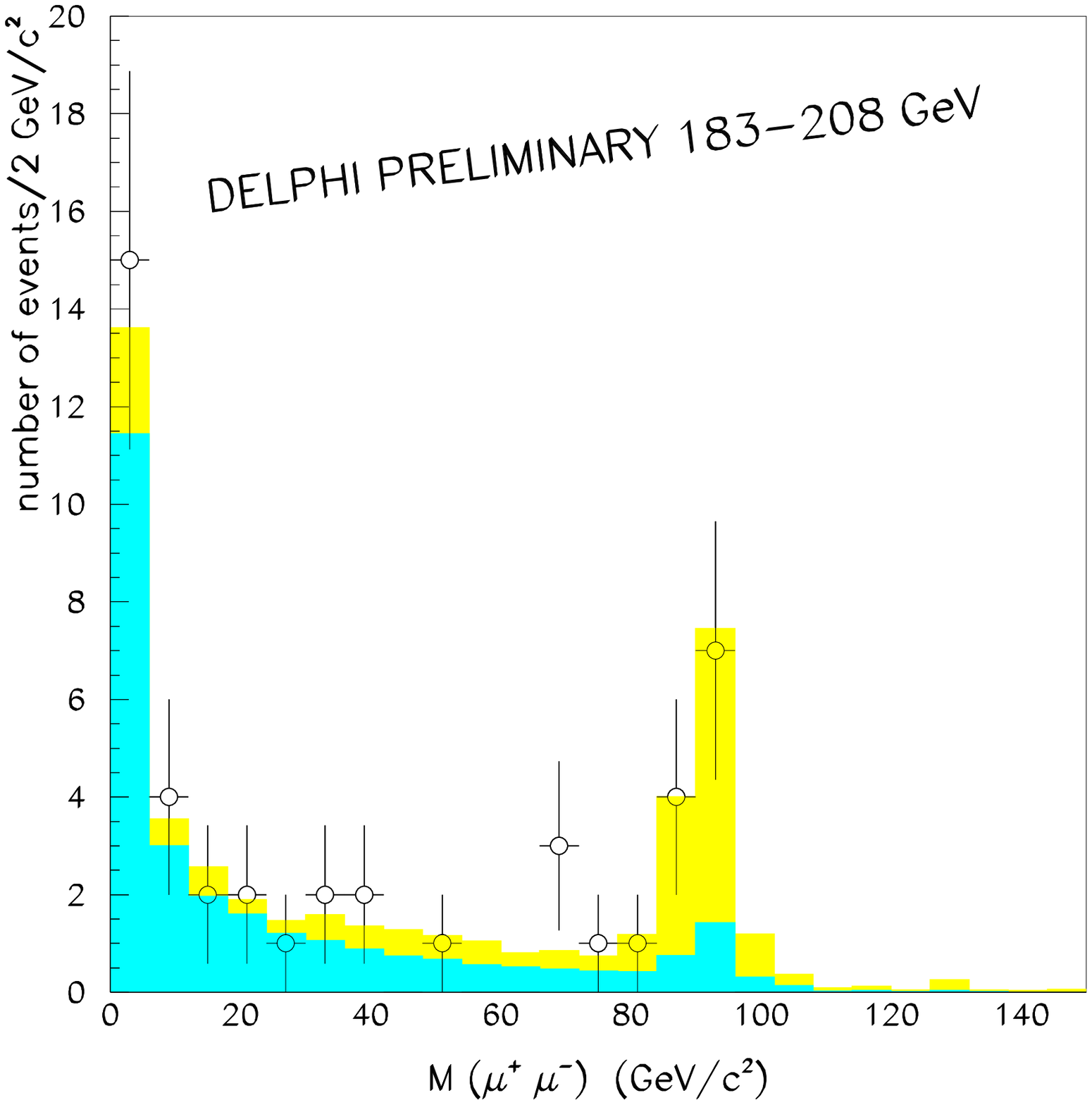,width=6.0cm} }
\vspace*{-3.8cm} 
\centerline{ \hspace*{12.1cm} ${\scriptscriptstyle \gamma^\star\rightarrow \mu^+\mu^-}$ \hspace*{0.7cm} ${\scriptscriptstyle Z\rightarrow \mu^+\mu^-}$ \hfill }
\vspace*{3.0cm} 
  \caption{On the left: distributions of the $ZZ$ probability in 4-jet channel for all the LEP2 data sample compared 
           with the SM predictions. Central figure: LEP combined results$^7$ 
           for $ZZ$ cross section.
           On the right: invariant mass distribution for 
           $\mu^+\mu^-$ pair in $q\bar{q}\mu^+\mu^-$ events.}
  \label{fig:zz} 
\vspace*{-0.1cm} 
\end{figure}

While $qqll$ events can be selected with high purity, the $qqqq$ and $qq\nu\nu$ channels are affected by a large amount of 
background. In these two channels, multivariate analyses are usually used to discriminate the signal from the background. In the 
four jets channel the $b$ tag techinque is used againts $WW$ events. 
The contamination is than mainly due to QCD events. 
This is clear in figure \ref{fig:zz} wich shows the output of the multivariate analysis developed by the DELPHI 
collaboration compared with the SM expectations for each process. 

The four experiments have measured the $ZZ$ cross section at all energies above 183~GeV. 
The combined LEP results are shown in figure \ref{fig:zz} and they are compared with the theoretical 
predictions calculated with the YFSZZ \cite{yfszz} and ZZTO \cite{yellow:4-ferm} programs.

The $ZZ$ production is sensitive to possible anomalous vertex interactions $ZZV$ ($V=Z/\gamma$). Following the 
parametrization suggested in \cite{hagiwara} there are four independent couplings: $f_i^V$ ($i=1,2$ and $V=Z/\gamma$)
that for $i=4$ are CP-odd and for $i=5$ are CP-even. The LEP combined results \cite{lwg:tgc} at the 95\% of confidence level are
$[-0.21,+0.23]$ for $f_4^\gamma$,  $[-0.40,+0.33]$ for $f_4^Z$, 
$[-0.37,+0.49]$ for $f_5^\gamma$ and $[-0.27,+0.29]$ for $f_5^Z$.  They are in agreement with the SM predictions $f_i^V=0$.
These measurements benefit of the higher energy run of the year 2000, only through the analyses performed by 
the ALEPH and DELPHI collaborations \cite{ntgc-f}. 

In the $Z\gamma^\star$ production the virtual photon decays into a pair of fermions wich are separated by a small angle.
The DELPHI collaboration \cite{zgstar} has presented the $Z\gamma^\star$ cross section for three different decay channels averaged 
over energies above 183~GeV:
$\sigma_{Z\gamma^\star\rightarrow qq\mu\mu} = (0.123 \pm 0.025)$~pb, 
$\sigma_{Z\gamma^\star\rightarrow qq\nu\nu} = (0.129 \pm 0.038)$~pb and
$\sigma_{Z\gamma^\star\rightarrow qqqq} = (0.074  \pm 0.045)$~pb. They are in agreement with the 
SM expectations: 0.098~pb for $qq\mu\mu$, $(0.092 \div 0.084)$~pb for $qq\nu\nu$ and 0.082~pb for qqqq.
\section*{Acknowledgments}
I thank my DELPHI colleagues working in these subjects for all the constructive 
discussions. 


\section*{References}
\begin{thebibliography}{99}
%
\bibitem{m01:mw} N. Watson, these proceedings.
\bibitem{yellow:lep2-1} {\it ``Report on the LEP2 workshop''}, CERN 96-01 (1996), vol.I.
\bibitem{zee-osaka} Carmen Palomares, proceeding of ICHEP 2000.
\bibitem{csww:00} ALEPH 2001-013 CONF 2001-010; 
                  DELPHI 2001-024 CONF 465; 
                  L3 Note 2638; 
                  OPAL Physics Note PN469.
\bibitem{racoonww}  A.~Denner, S.~Dittmaier, M.~Roth and D.~Wackeroth, {\it Nucl. Phys.} B {\bf 587} (2000) 67.
\bibitem{yfsww}     S. Jadach {\it et al.}, {\it Phys. Lett.} B {\bf 417} (1998) 326, hep-ph/0103163 
\bibitem{ewg:cs} The LEP collaborations and the LEP $WW$ Working Group, LEPEWWG/XSEC/2001-01 \\
                 http://lepewwg.web.cern.ch/LEPEWWG/lepww/4f/Winter01/xsec$_-$moriond01.ps.gz .
\bibitem{pdg}     Particle Data Group, D.E. Groom {\it et al.}, {\it Eur. Phys. J.} C {\bf 15} (2000) 1.
\bibitem{tgc-aleph} ALEPH 2001-027 CONF 2001-021.
\bibitem{lwg:tgc} The LEP collaborations and the LEP TGC Working Group, LEPEWWG/TGC/2001-01 \\ 
                   http://lepewwg.web.cern.ch/LEPEWWG/lepww/tgc/note.ps.gz . 
\bibitem{wpol-l3} L3 Note 2636.
\bibitem{wen-aleph} ALEPH 2001-017 CONF 2001-014.
\bibitem{yfszz}   S. Jadach, W. Placzek, B.F.L. Ward, {\it Phys. Rev.} D {\bf 56} (1997) 6939.
\bibitem{yellow:4-ferm} {\it ``Reports of the working groups on precision calculations for LEP2 Physics''}, CERN 2000-009 (2000).
\bibitem{cszz}  ALEPH 2001-006 CONF 2001-003; 
                DELPHI 2001-015 CONF 456; 
                L3 Note 2641; 
                OPAL Physics Note PN469.
\bibitem{hagiwara} K. Hagiwara, K. Hikasa, R.D. Peccei and D. Zeppenfeld, {\it Nucl. Phys.} B {\bf 282} (1987) 253.
\bibitem{ntgc-f} ALEPH 2001-014 CONF 2001-011; DELPHI 2001-014 CONF 455. 
\bibitem{zgstar} DELPHI 2001-008 CONF 449.
\end{thebibliography}
\end{document}